\documentclass[11pt]{article}

\usepackage{bm}
\usepackage{color}

\usepackage{amsmath}
\usepackage{amssymb}
\usepackage{graphicx}
\usepackage{amssymb}
\usepackage{epstopdf}

\newcommand \beq{\begin{eqnarray}}
\newcommand \eeq{\end{eqnarray}}

\def\k{{\boldsymbol k}}

\def\v{{\boldsymbol v}}

\begin{document}

\title{EMMI Rapid Reaction Task Force on\\
{\bf``Thermalization in Non-abelian Plasmas''}}
\author{J\" urgen Berges$^1$, Jean-Paul Blaizot, Fran\c cois Gelis$^2$\\[1ex]
$^1$ Universit\"at Heidelberg, Institut f{\"u}r Theoretische Physik,\\
Philosophenweg 16, 69120~Heidelberg, Germany\\
and ExtreMe Matter Institute EMMI,\\
GSI Helmholtzzentrum f\"ur Schwerionenforschung GmbH,\\ 
Planckstra\ss e~1, 64291~Darmstadt, Germany\\[1ex]
$^2$ Institut de Physique Th\'eorique (URA 2306 du CNRS),\\ 
CEA/DSM/Saclay, 91191, Gif-sur-Yvette Cedex, France
}
\date{}

\maketitle

\begin{abstract} {Recently, different proposals have been put forward
    on how thermalization proceeds in heavy-ion collisions in the
    idealized limit of very large nuclei at sufficiently high
    energy. Important aspects of the parametric estimates at weak
    coupling may be tested using well-established
    classical-statistical lattice simulations of the
    far-from-equilibrium gluon dynamics. This has to be confronted
    with strong coupling scenarios in related theories based on
    gauge-string dualities. Furthermore, closely related questions
    about far-from-equilibrium dynamics arise  in early-universe cosmology
    and in non-relativistic systems of ultracold atoms. These were central topics of the EMMI Rapid Reaction
    Task Force meeting held on December 12-14, 2011, at the University
    of Heidelberg, which we report on.  }
\end{abstract}

\section{Motivation}

The topic of ``Thermalization in expanding Non-abelian Plasmas'' is one
of the most pressing issues in the physics of \emph{ultra-relativistic
  heavy ion collisions}. Heavy ion experiments involve at least initially
far-from-equilibrium systems of strongly interacting matter described
by quantum chromodynamics (QCD). The data suggest a \emph{rapid
  apparent thermalization} of the produced matter, with robust
collective phenomena, whose theoretical understanding from QCD
still represents a great challenge. Non-abelian plasmas out of equilibrium
are also important in the {\it early universe}. A central question
there is the reheating after inflation, a period of strongly
accelerated expansion which dramatically dilutes all matter and
radiation. The subsequent rapid transition to a thermal Hot-Big-Bang
cosmology seems to be crucial for our current understanding of the
evolution of the universe.

For a sufficiently \emph{weakly coupled plasma} \emph{close to
  equilibrium} much pro\-gress has been achieved in recent years. The
dynamics in this case may be efficiently described in terms of kinetic
theory including elastic and inelastic processes. However, much less
is known about the relevant \emph{dynamics far from equilibrium}. This
concerns even the (a priori) theoretically clean case of very large
nuclei at sufficiently high energy. In this idealized limit, the
matter formed shortly after the collision is believed to be described
by the ``color glass condensate'', with a characteristic
``saturation'' momentum scale $Q_s$ that grows with the energy and the
size of nuclei. When $Q_s$ is large, the coupling constant is
small. Yet the gluons at saturation appear strongly coupled because
their low momentum modes have \emph{very high occupancy}.

Recently, different groups have put forward \emph{scenarios of how
  thermalization proceeds} in this framework. In contrast to earlier
works, there are now detailed proposals covering the dynamics from
immediately after the collision until well after the plasma becomes
nearly equilibrated. \emph{Plasma instabilities}, which have been
extensively studied in recent years, seem to dominate the early-time
dynamics. Initially they drive the system close to \emph{isotropy}
with a gluon density that is parametrically large when compared to a
system in thermal equilibrium with the same energy density. Different
scenarios of what happens next have been proposed. In one scenario, it
is argued that Bjorken expansion is strong enough to prevent the
system to become isotropic on short time scales. In another scenario,
it is argued that scattering, while probably not sufficient to fully
restore isotropy, may nevertheless maintain the system in a state of
fixed anisotropy for a long time. For instance, if particle number
conserving processes dominate, the over-population of low-momentum
modes may lead to the dynamical generation of a \emph{Bose-Einstein
  condensate}, corresponding to a large occupation of the zero
momentum mode. This may be accompanied or preceded by the formation of
\emph{turbulent cascades} at intermediate stages. Since the
\emph{parametric estimates} that have been presented allow for
different types of solutions, it is crucial to test the analytic
predictions against \emph{lattice simulations} of the
far-from-equilibrium gluon dynamics. Here it is important that the
dynamics is, to a large extent, \emph{classical}, so that
well-established classical-statistical simulation techniques can be
employed.

The special situation of having very detailed analytical proposals and
the prospect of testing them with state-of-the-art simulation
techniques led us to expect that a significant advance can be achieved
by bringing together leading scientists working on this very specific
set of questions. Therefore, a particular emphasis was to confront
weak-coupling parametric estimates with classical lattice gauge theory
results for the far-from-equilibrium dynamics. This had to be compared
to strong coupling scenarios in related theories based on gauge-string
duality. Furthermore, closely related questions about
far-from-equilibrium dynamics in early-universe cosmology and in
non-relativistic systems of ultracold atoms were taken into
account. This very focused meeting was organized as the first EMMI Rapid Reaction Task Force, 
a new instrument of the ExtreMe Matter Institute. The agenda of the
meeting and the list of participants are available from the website
http://www-aix.gsi.de/conferences/emmi/tnp2011/.

\section{Thermalization of the quark-gluon plasma}

\subsection{Phenomenological perspectives}

As recalled in the previous section, the problem of thermalization in
ultra-relativistic heavy ion collisions proceeds from important and
robust empirical evidence, obtained both from RHIC and LHC \cite{EXP}:

- Matter produced in heavy ion collisions exhibits fluid behavior from
very early time on (collective flow is visible in several moments of
the azimuthal distributions, and is sensitive to fluctuations of the
initial geometry);

- The fluid has very special transport properties, in particular a
small value of the shear viscosity to entropy density ratio, $\eta/s$.
            
These results are usually interpreted in terms of (viscous)
hydrodynamics, but this raises at least two questions. Fluid-like
behavior requires some degree of local equilibration or at least
isotropization. How is this achieved? The small value of $\eta/s$
points to strong interactions (short mean free path). What is the
origin of these strong interactions in a system whose dynamics is
governed by a coupling constant that is not very large? We shall get
back to this second issue below. First we focus on the issue of local
equilibrium.

Note that what the data tell us is not completely unambiguous. Thus,
as was emphasized by M.\ Strickland \cite{Martinez:2010sd} during the
workshop, and as is well known to some hydro practitioners (see
e.g.~\cite{Florkowski:2010in}), collective flow measurements provide in
fact little evidence (if any) that the longitudinal pressure is
identical to the transverse pressure. Let us recall that the energy
momentum tensor that enters the hydrodynamical equations,
$\partial_{\mu}T^{\mu\nu}=0$, has, in the local rest frame of the
fluid, the generic form $T^{\mu\nu}={\rm diag}(\epsilon,
P_T,P_T,P_L)$, where $P_T$ and $P_L$ are the transverse and the
longitudinal pressure, respectively, and $\epsilon$ is the energy
density.  For massless partons, $T^\mu_{\,\mu}=0$ (up to corrections
due to the trace anomaly, that are sizable only near $T_c$), and
$\epsilon=2P_T+P_L$.  In case of complete local equilibrium,
$P_T=P_L$, and $\epsilon=3P$. However, complete local equilibrium is
not necessary to get hydrodynamic behavior. This was in particular
demonstrated by R.\ Janik \cite{Heller:2011ju} who showed, within the
AdS/CFT framework, that viscous hydro starts to be valid already at
times where $P_L/P_\perp\sim 0.5$, that is well before
$P_L/P_\perp\sim 1$.

The main message here is that we should perhaps not conclude too
hastily that the data provide clear evidence for local thermal
equilibrium. Some significant anisotropy of the momentum distributions
(i.e., incomplete local equilibrium) may still be compatible with
viscous hydrodynamic behavior. The issue of what is meant by
\emph{thermalization time } should perhaps be re-examined within this
context.  Furthermore, an interesting open question is whether there
are observables sensitive to a difference between the longitudinal and
the transverse pressures, $P_L $ and $P_T$?

\subsection{Weak or strong coupling?}

We turn now to the second empirical observation, namely the apparent
smallness of $\eta/s$. This is taken as evidence for short mean free
path and strong coupling. Taken to the extreme, this observation
motivates the infinite coupling approach based on the AdS/CFT
correspondence.  This approach, reviewed by R. Janik, provides a
powerful framework to derive successive viscous hydrodynamical
approximations, and specify their regimes of validity. As already
mentioned, a particularly striking result of such calculations is the
indication that viscous hydrodynamics starts to be valid well before
the pressures are equilibrated.

However, while AdS/CFT provides a very elegant framework for
describing the collective, fluid-like behavior of the quark-gluon
plasma, it does not answer the question of how this behavior emerges
from a system of quarks and gluons whose elementary interactions are
controlled by a small coupling constant. In the regimes that we are
considering the strong coupling constant is not huge, typically
$\alpha_s\sim 0.3 - 0.4$ at RHIC or LHC energy. Thus, although the
small value of $\eta/s$ argues against the validity of strict
perturbation theory, it would be surprising if weak coupling
approaches would be completely misleading. For instance, such weak
coupling techniques (involving appropriate resummations) provide a
quantitative understanding of the thermodynamics of lattice QCD for
temperatures higher than 2 to 3 $T_c$. Perhaps more to the point, our
present understanding of the initial stages of heavy ion collisions is
based on weak coupling (for large nuclei and large energies). The
initial gluon distribution is characterized by the saturation momentum
$Q_s$, which grows as the energy increases and as the size of the
nuclei grows. In the limit of large energies and large nuclei, $Q_s$
becomes large, and since this is the scale that controls the running
coupling, $\alpha_s$ is small. However, even in this weak-coupling
case the gluons at saturation appear strongly coupled because they
reach high occupancies $\sim 1/\alpha_s$.

There are therefore strong motivations to pursue efforts to understand
the thermalization of the quark-gluon plasma from weak-coupling
considerations. The issues discussed in the remaining part of this
report are to be understood within this context.

\subsection{Instabilities and initial conditions} 

It has been recognized that the color fields that could be present in
the initial stages of nucleus-nucleus collisions may exhibit
instabilities that can play an important role in the early dynamics
\cite{Mrowczynski:2008jq,Mrowczynski:2009aq}. As
discussed by G.~Moore, these instabilities, which can be of various
kinds (Weibel instability, Nielsen-Olesen instability, etc.) involve
primarily color magnetic fields, and their role is mainly to
redistribute the momentum directions, with no change in the
energy. These instabilities can isotropize the momentum distributions
much more rapidly than ordinary scattering processes at sufficiently
weak coupling.

Quantitative studies of the evolution at early times are typically performed either in simulations of semi-classical transport approaches such as the hard-loop framework~\cite{HL2} and taking into account backreactions on the momentum distribution of the hard particles~\cite{HL3} or in numerical simulations of the classical-statistical field theory \cite{Romatschke:2005pm,Berges:2007re,Fukushima:2011nq}. Instabilities are also closely related to entropy
production in the system, as was emphasized by B.~Muller and
A.~Schafer \cite{Muller:2011ra,Kunihiro:2010tg}.

Whether momentum distributions remain isotropic or not in the presence
of the strong longitudinal expansion is an important open question. In
fact it was recently pointed out that, within a weak-coupling
description, two scenarios could be possible
\cite{Kurkela:2011ub,Kurkela:2011ti}. One in which isotropy could be
maintained throughout the expansion, i.e., the anisotropy remains at
most of order one. Another scenario in which the longitudinal
expansion is so strong as to keep the momentum distributions
anisotropic until the very late stages where thermalization occurs. In
order to see whether in the idealized world of truly weak coupling,
one scenario prevails over the other, it is necessary to push the
analysis beyond the simple parametric estimates. This requires not
only to get the power dependence of various observables on the strong
coupling $\alpha_s$, but also to get the \emph{coefficients multiplying}
$\alpha_s^n$. This splits into two questions: one is whether one can do
this by making a careful separation of all scales in the weak-coupling
limit (similar to the philosophy leading to hard-loop effective
theories~\cite{HL2}); another is whether the same goal can maybe better achieved with
methods that treat physics at more than one scale at the same time
(e.g.\ classical-statistical lattice gauge theory~\cite{Romatschke:2005pm,Berges:2007re,Fukushima:2011nq}, or particle plus soft-field simulations~\cite{HL3}).  Note finally that the weak-coupling analysis
of instabilities may be complicated by effects due to the
overpopulation of the initial gluon distribution, while it can well be described using classical fields on the lattice, as we shall discuss below.

The physics of instabilities is influenced by the \emph{initial
  conditions}. In particular, these determine the initial size of
boost non-invariant fluctuations which are further amplified by the
instability. At some point they could grow large enough for non-linear
corrections to become important. These non-linearities can lead to a
secondary stage of enhanced amplification of modes in a momentum
region exceeding by far the initially unstable band. Nonlinear amplifications 
with strongly enhanced growth rates have been identified in fixed-box studies 
of plasma instabilities in classical Yang-Mills theory \cite{Berges:2007re,Fujii:2009kb}. 
Secondary instabilities have also been analyzed in longitudinally expanding systems for scalar field theories, 
where the evolution of the quantum theory can also be studied \cite{Berges:2012iw}. 
They could significantly speed-up the isotropization process, but their relevance depends to some
extent on the initial conditions for these modes. All this leads to
the important question of whether one can perform a first-principle
calculation of the initial conditions that prevail at the beginning of
heavy ion collisions. This is what the glasma picture is claiming to
achieve as discussed by R.\ Venugopalan \cite{Dusling:2011rz}: strong
magnetic and electric fields are produced from highly localized
sources, with the fields subsequently decaying into particles on a
time scale of order $1/Q_s$.

However, one could envisage a purely partonic description, without
fields, with parton-parton collisions putting gluons on-shell in a
time scale of order $1/Q_s$. It would be very useful to clarify
further to which degree one controls the space-time evolution of the
system at the very early stages. The evolution of such a partonic
system using transport theory has been discussed 
by C.~Greiner \cite{El:2009zz,Reining:2011xn}.

\subsection{Turbulence and cascade towards the infrared}

Related to the physics of instabilities is that connected with
turbulence and various types of cascades in momentum space
\cite{Arnold:2005ef,Arnold:2005qs}. These phenomena can be studied
using numerical simulations of classical fields, with suitable
averages over the initial conditions, as was explained by
J.~Berges \cite{Berges:2008mr,Fukushima:2011nq}. Let us recall that such classical field simulations can
accurately describe the regimes of large occupation ($1\ll n(p)$) all
the way up to the overpopulated regime ($n(p)\sim 1/\alpha_s$). They
fail of course when occupations become smaller than unity, at which
point quantum effects need to be taken into account. These regimes of
large occupations can also be described by kinetic theory (whose
validity extend toward the quantum regime), except a priori in the
overpopulated regime where multiparticle processes become all of the
same order of magnitude. There are, however, cases where one can resum
these high order processes into an effective kinetic theory with only
$2\to 2$ processes and effective (momentum dependent) vertices. In
this particular case, an effective kinetic theory can also be used to
describe the overpopulated regime.

In the case of scalar theory, detailed calculations can be performed
in these various regimes, using both classical field simulations, as
well as large-$N$ 2PI techniques (that allow for a quantum
treatment) \cite{Berges:2012us}. What emerges from these calculations is that the dynamics
is dominated by a dual cascade, one towards the infrared, associated
to the approximate conservation of particle number, the other towards
the ultraviolet, associated with approximate conservation of the
energy. For overpopulated initial conditions, a Bose condensate
develops and remains present in the system until eventually inelastic
scattering depletes it. Exponents associated with
these two cascades are under analytical control and are well
reproduced in the simulations. In particular, the value of the
exponent for the UV cascade can be understood as arising from the
existence of the condensate leading to an effective cubic interaction.

The extension of these results to the case of non-abelian gauge
theories offers a number of challenges that were very much
discussed. It was emphasized by A.~Mueller \cite{Mueller:2006up} that
cascades can be very different in scalar theories and in gauge
theories. In scalar theories, the fact that the interactions are
nearly local in momentum space typically plays a crucial role to
explain the flow of momenta. This is not the case in gauge
theories (where we can go from hard to soft momenta in one
collision). In the case of scalar theories, one can control to a large
extent analytically what is going on. In the case of gauge theories,
the rigorous determination of the scaling exponents is complicated by
questions of convergence of the relevant integrals whose scaling behavior one wants
to study. Remarkably, classical Yang-Mills theory simulations starting from over-populated initial conditions indicate the same weak wave turbulence exponent as for scalars \cite{Berges:2012us}. This is also what one gets from an analytical determination of turbulent scaling exponents ignoring questions of convergence.
Although the present gauge theory results exhibit behavior reminiscent 
of the scalar case, more work
is needed in order to get conclusive results, in particular, about the infrared behavior. Finally similar
approximations as the large-$N$ 2PI expansion, which give analytical
control for the scalar case, are much more difficult to implement in
gauge theories.

It should also be kept in mind that most simulations so far ignore the
longitudinal expansion, which is an essential aspect of the dynamics
of heavy ion collisions.

\subsection{Bose-Einstein condensation}

The standard picture of heavy ion collisions assumes that the gluons
that contribute dominantly to the energy density of the produced
matter are freed over a time scale of order $\tau_0\sim Q_{\rm
  s}^{-1}$, and have typical transverse momenta of order $Q_{\rm
  s}$. Their contribution to the energy density is $ \epsilon_0\sim
{Q_{\rm s}^4}/{\alpha_{\rm s}}\,, $ while their number density is $
n_0\sim {Q_{\rm s}^3}/{\alpha_{\rm s}}.  $ One may characterize this
initial distribution of gluons by the dimensionless combination
$n_0\;\epsilon_0^{-3/4}\sim 1/\alpha_{\rm s}^{1/4}$.  In comparison,
in an equilibrated system of gluons at temperature $T$, $
\epsilon_{\rm eq} \sim T^4$, $ n_{\rm eq}\sim T^3$, so that $ n_{\rm
  eq}\epsilon_{\rm eq}^{-3/4}\sim 1$.  There is therefore a mismatch,
by a large factor $\alpha_{\rm s}^{-1/4}\gg 1$ (in weak coupling
asymptotics, $\alpha_{\rm s} \ll 1$), between the value of
$n\epsilon^{-3/4}$ in the initial condition and that in an
equilibrated system of gluons.  This mismatch can be interpreted as an
``overpopulation'' of the initial distribution.

Gluons in a plasma become ``massive'' as a result of their
interactions, and the number of massive gluons can be controlled by a
chemical potential. It is easy to show, however, that a chemical
potential will not help decreasing the overpopulation if the system is
driven to equilibrium by {\it elastic collisions alone}.  In this
case, Bose condensation will occur, leading to an equilibrium state in
which most of the particles are to be found in the condensate, while
most of the energy density remains carried by thermal particles. Note
that inelastic, particle number changing processes preclude the
possibility that the true equilibrium state be a Bose condensate, but
it leaves open the possibility that a transient condensate develops
during the evolution of the system.

The question of how the system evolves towards its equilibrium state
was addressed in \cite{Blaizot:2011xf}, using a simple kinetic
equation of the form $(\partial_t +\v\cdot \nabla )f(\k,X) =C_\k[f]\;
,$ where $C_k[f]$ is the usual collision integral for 2 to 2 processes
(mean field contributions are left out in the left hand-side: such
terms are responsible for the instabilities discussed earlier, and it
was just assumed in \cite{Blaizot:2011xf} that their main role is to
maintain some degree of isotropy). One finds then, in the small angle
approximation for the collision integral, that the overpopulated
plasma is driven towards Bose condensation. Remarkably, in the regime
where $f\gg 1$ ($f\sim 1/\alpha_{\rm s}$), all dependence on
$\alpha_{\rm s}$ disappears, so that the system seems strongly coupled
(a property that is also manifest in the classical field
description). Inelastic particle production or annihilation processes
modify the collision integral, but in the overpopulated regime, the
inelastic scattering does not change the basic time scales in the
problem. Whether the condensate forms or not can then only be answered
after a detailed numerical analysis of the solution of the transport
equation. Here it would be very useful to implement suitable
resummations to extend the validity of the transport equation to the
overpopulated regime, similar to what is achieved with large-$N$ 2PI
techniques for the scalar case.

The dynamical formation of a condensate was demonstrated, in the case
of scalar field theories, by J.~Berges and D.~Sexty \cite{Berges:2012us} who showed that condensation occurs as a consequence of an inverse particle cascade with a universal power-law spectrum. This particle transport towards low momenta is part of a dual cascade, in which energy is also transfered by weak wave turbulence towards higher momenta as mentioned above. Condensation was also discussed by T.~Epelbaum et al.~\cite{Dusling:2010rm,Epelbaum:2011pc} who presented
results of their simulations. Further evidence was given in the
talks by I.~Tkachev \cite{Micha:2004bv} on the early universe based on the value of the
exponent for the UV cascade, and by T.~Gasenzer 
\cite{Nowak:2011sk} in the context of cold atoms emphasizing the role of non-trivial topological configurations for the dynamics. Whereas a Bose condensate
develops and remains present in the relativistic system until eventually inelastic
scattering depletes it, for the non-relativistic theory with conserved particle number no decay of the condensate due to number changing processes is observed \cite{Berges:2012us}.

Whether condensation can occur in non-abelian gauge theories is an
issue that was very much debated. One may indeed wonder whether the
notion of a condensate makes sense in the context of non-abelian gauge
theories, as this is seemingly in conflict with gauge
invariance. However, it was pointed out that, in analogy with the
parton model (which, strictly speaking, only makes sense in the Light
Cone gauge), Bose condensation could be a phenomenon that becomes
manifest only in a specific gauge (Coulomb gauge?). In other gauges,
the description of the effect of the large population of low momentum
modes might be more subtle. It was pointed out, in view of past
experience with pion condensation, that it would be important to
estimate the size of the domains where condensation could take place,
since if these domains are too small, the phenomenon would have very
little impact on the dynamics. The role of topological defects, which
play an important role in cosmology and in the dynamics of cold atoms
was also discussed. Overall, the topic has triggered many interesting
discussions, which may be expected to lead to further fruitful
studies.

\section{Conclusions}

The particular workshop format as a Rapid Reaction Task Force of the ExtreMe Matter Institute EMMI allowed for in-depth discussions of many important issues related to the far-from-equilibrium behavior of non-abelian gauge theories: plasma instabilities, various cascading phenomena, the amplification of low momentum modes leading possibly to Bose-Einstein condensation, etc. Classical-statistical lattice gauge simulations appear as an efficient tool to explore some of these questions, while analytical methods should be able to handle the weak coupling regimes, in particular the regime of large occupation numbers where  strong correlations can appear. This could be confronted with the impressive progress in the understanding of the dynamics of non-trivial field theories in the infinitely strong coupling limit based on gauge-string dualities. 

 It is important to work out further the relevant differences and similarities between gauge and scalar degrees of freedom out of equilibrium. It is, for instance, remarkable that classical simulations starting from over-populated initial conditions indicate for Yang-Mills theory the same turbulence exponents as for scalars. The question of how to generalize the notion of Bose condensation to gauge theories and to devise suitable gauge-invariant measures of condensation is a non trivial task.  Furthermore, closely related questions about far-from-equilibrium dynamics in early-universe cosmology and in non-relativistic systems of ultra-cold atoms give valuable insight and point to universal aspects of far-from-equilibrium phenomena.

Many issues remain to be explored before we have a realistic scenario of how thermalization proceeds in heavy ion collisions. Some of the phenomena expected from the weak coupling analysis may involve too long time scales to be directly relevant. The role of the longitudinal expansion needs to be further clarified: This should allow one to distinguish between the scenario where isotropy is maintained throughout the expansion, and the situation in which the longitudinal expansion is so strong as to keep the momentum distributions anisotropic until the very late stages where complete thermalization occurs. There remains also the  important question of whether one can perform a first-principle calculation of the conditions that prevail at the beginning of
heavy ion collisions, as in the glasma picture.\\

\noindent
{\Large\bf Acknowledgments}\\

\noindent
This work was supported by the Alliance Program of the Helmholtz
Association (HA216/EMMI). We thank all the participants to this
meeting for having created a very stimulating atmosphere and for the
lively discussions that this report attempts to reflect.



\providecommand{\href}[2]{#2}\begingroup\raggedright\endgroup

\end{document}